\RequirePackage{fix-cm}
\documentclass[smallextended]{svjour3-rev}       
\smartqed  
\usepackage{graphicx}
\usepackage{color}
\begin{document}

\title{Surface tension and Laplace pressure in triangulated surface models for membranes without fixed boundary
}

\titlerunning{Surface tension and Laplace pressure in triangulated surface models}        

\author{Hiroshi Koibuchi\and
        Andrey Shobukhov\and 
        Hideo Sekino
}


\institute{Hiroshi Koibuchi \at
  Department of Mechanical and Systems Engineering, 
  National Institute of technology, Ibaraki College, 
  Nakane 866, Hitachinaka, Ibaraki 312-8508, Japan\\
              \email{koibuchi@mech.ibaraki-ct.ac.jp}           
           \and
            Andrey Shobukhov \at
 Faculty of Computational Mathematics and Cybernetics, Lomonosov Moscow State University, 
119991, Moscow, Leninskiye Gory, MSU, 2-nd Educational Building, Russia
           \and 
           Hideo Sekino \at
Computer Science and Engineerings, Toyohashi University of Technology, 
Hibarigaoka 1-1, Tenpakuchou, Toyohashi, Aichi 441-8580, Japan
}

\date{}

\maketitle

\begin{abstract}
A Monte Carlo (MC) study is performed to evaluate the surface tension $\gamma $ of spherical membranes that may be regarded as the models of the lipid layers. We use the canonical surface model defined on the self-avoiding triangulated lattices. The surface tension $\gamma $ is calculated by keeping the total surface area $A$ constant during the MC simulations. In the evaluation of $\gamma $, we use $A$ instead of the projected area $A_p$, which is unknown due to the fluctuation of the spherical surface without boundary. The pressure difference ${\it\Delta}p $ between the inner and the outer sides of the surface is also calculated by maintaining the enclosed volume constant. Using ${\it\Delta}p $ and the Laplace formula, we obtain the tension, which is considered to be equal to the frame tension $\tau$ conjugate to $A_p$, and check whether or not $\gamma $ is consistent with $\tau$. We find reasonable consistency between $\gamma$ and $\tau$ in the region of sufficiently large bending rigidity $\kappa$ or sufficiently large $A/N$. It is also found that $\tau$ becomes constant in the limit of $A/N\to \infty$ both in the tethered and fluid surfaces.
\end{abstract}

\section{Introduction}
Contemporary chemical applications of surface models are closely related to the study of the lipid bilayers. Initially the lipid membranes were described as a two-dimensional fluids with the protein molecules diffusing in them \cite{Scott-BPJ-1991,Paster-COSB-1994}. But the observed variety of the lipid structures in cells \cite{Merz-COSB-1997,Tobias-COSB-1997,Nagle-COSB-2000} proved that the lipid molecules were much more than just a solvent for proteins. Since then the systems consisting of such molecules were extensively studied using both experimental and computational methods. It should be mentioned that in spite of the early Monte-Carlo approach in \cite{{Scott-BPJ-1991},Paster-COSB-1994}, later the researchers mainly used the molecular dynamics (MD) technique \cite{Merz-COSB-1997,Tobias-COSB-1997,Venable-JCP-2000,Chiu-BJ-1995}. Particular attention was paid to the relation between the structural and dynamical properties of the membranes and their biochemical functions inside the cell. The MD method was applied for simulation of lipid bilayers \cite{Venable-JCP-2000,Chiu-BJ-1995,Chiu-JCC-1999,Feller-BJ-1996,Lindahl-JCP-2000,Marrink-JPCB-2001,Lindahl-BJ-2000}. It gave the first glance at the membrane structural features at the molecular level. But at the same time it proved to be capable of modeling not more than several nanoseconds of evolution for a realistic molecular structure in a reasonable computational period: the simulation time range from 10 to 60 nanoseconds was described in \cite{Lindahl-BJ-2000} as a serious achievement. For this reason in  \cite{Chiu-JCC-1999} even a combination of the MD and MC approaches was proposed.

In the present paper we pay special attention to the final steady configurations of the investigated membranes, and thus we apply the MC technique. We study the mechanical characteristics of membranes, such as the bending modulus, the surface tension $\gamma$ and the pressure difference ${\it\Delta}p \!=\!p_{\rm in}\!-\!p_{\rm out}$, where $p_{\rm in} (p_{\rm out})$ is the pressure inside (outside) the surface.  These physical quantities, previously studied in \cite{Scott-BPJ-1991,Paster-COSB-1994,Merz-COSB-1997,Tobias-COSB-1997,Nagle-COSB-2000,Venable-JCP-2000,Chiu-BJ-1995,Chiu-JCC-1999,Cai-Lub-PNelson-JFrance1994,Ambjorn-PRL1987,Ambjorn-NPB1993,WHEATER-JP1994,Dobreiner-et-al-PRL2003,Fournier-PRL2008,Imparato-JCP2006,PDPJB-EPJE2004},  potentially reflect the microscopic interactions of the constituent molecules. 

Experimentally the bending modulus and the spontaneous curvature of a fluid membrane were obtained using the so-called flicker spectroscopy technique, in which numerically generated membrane shapes are fitted to experimental ones \cite{Dobreiner-et-al-PRL2003}. The same method called the contour detection technique was applied to experimentally measured shapes of membranes for extracting the bending modulus and the surface tension \cite{PDPJB-EPJE2004}.   

Therefore, it is interesting to study $\gamma$ and ${\it \Delta}p$ by means of stochastic simulation of mechanical processes on the triangulated surface models, such as the Helfrich-Polyakov (HP) model \cite{HELFRICH-1973,POLYAKOV-NPB1986,NELSON-SMMS2004}. The frame tension $\tau$ is defined via the macroscopic surface energy, which equals $\tau A_p$, where $A_p$ is called the projected area \cite{David-Leibler-JPF1991}. This $A_p$ may be regarded as the area contained within the boundary $\Gamma (\subset\! {\bf R}^2)$ fixed in ${\bf R}^3$, while the real surface area is denoted by $A$. In the case of discrete HP model, $A$ is given by the sum of triangle (or microscopic) areas. We use the term "frame (surface) tension $\tau$ ($\gamma$)" for the surface tension conjugate to $A_p$ ($A$) hence force \cite {Fournier-PRL2008,David-Leibler-JPF1991}. It is widely accepted that on the surfaces spanning $\Gamma$ the surface tension $\gamma$ is in general different from the frame tension $\tau$ \cite{Fournier-PRL2008}. To the contrary, these $\gamma$ and $\tau$ are expected to be identical when the surface is sufficiently smooth \cite{Fournier-PRL2008} and there is no difference between $A$ and $A_p$.

In the experimental measurements, the projected area $A_p$ is also used to obtain the frame tension $\tau$ in a variant of the Laplace formula for the cell deformed between the parallel plates \cite{Foty-etal-PRL1994,Foty-etal-Devlop1996}. The pressure difference ${\it\Delta}p$ is obtained by measuring both the contact area at the plates and the force $F$ on the plates, and then the frame tension $\tau$ may be found using the Laplace formula. Thus, the boundary condition imposed on the cell by the plates fixes the shape parameters such as the contact area and the radius, which play the role of $A_p$ for the spherical membrane.

However, $A_p$ is not always well-defined on the surfaces without boundary.  Therefore, the frame tension cannot be obtained directly on the sphere without the fixed boundary. Nevertheless,  the pressure difference ${\it\Delta}p$ can be obtained, and using the ${\it\Delta}p$ and the Laplace formula,  we are able to find the effective frame tension $\gamma|_{V,{\rm Lap}}$.

The purpose of the present study is to find whether or not $\gamma$, $\gamma|_{V,{\rm Lap}}$ and the simulation techniques are well defined for spherical membranes without the fixed boundary. For this purpose,  in the present paper we report the MC data including those in  \cite{KSS-Msquare-2014}  with the detailed information on the calculation formula in a self-contained manner. During the MC sweeps (MCS), $A$ is kept fixed at the constant value $A_0$ (within the error $0.35\%$ for example). Almost the same constraint is imposed on the enclosed volume to obtain the pressure difference ${\it\Delta}p$ in the constant volume simulations. From ${\it\Delta}p$ and the Laplace formula, we obtain the tension $\gamma|_{V,{\rm Lap}}$, which can be regarded as the frame tension $\tau$ as mentioned above.  One additional point that should be noted is as follows: The area/volume constant MC simulation (ACMCS/VCMCS) is mathematically considered to be connected with the mapping ${\bf r}:M\to {\bf R}^3$ under the constraint of constant area/volume, where $M$ is a two-dimensional surface in the context of Polyakov's model for strings \cite{POLYAKOV-NPB1986}. These two mappings are in general considered to be very different from each other, because the surface morphology with the constant area is different from that with the constant volume. Thus, it is interesting to see a relation between the surface model and these constrained mappings in the simulations results for  $\gamma$ and $\gamma|_{V,{\rm Lap}}$.

It must also be noted on a relation between the fluid surface model defined on dynamically triangulated surfaces and the liquid phase of the lipid membranes. The lipid molecules undergo a first order transition, which is the so-called main transition, between the gel and liquid phases. The free diffusion of lipids seen in the liquid phase is greatly reduced in the gel phase. On the other hand, the triangulated lattice models can also be divided into two groups; one is the fixed connectivity model and the other is the dynamically triangulated model. The latter one is called the fluid surface model, because the vertices (of triangles) can diffuse freely over the surface due to the dynamical triangulation. On the fixed-connectivity lattices, the vertices can fluctuate only locally and hence have no free diffusion. Therefore, the model in this paper, which is a fluid model, corresponds to membranes in the liquid phase at least. 

\section{Models}\label{model}
\subsection{Surface tension $\gamma$}
The discrete Hamiltonian of the model is defined on a triangulated sphere with the vertex position ${\bf r}_i(i\!=\!1,\cdots,N)$ and the triangulation ${\mathcal T} $ \cite{Ambjorn-NPB1993} such that 
\begin{eqnarray}
\label{Disc-Eneg-tension} 
&& S({\bf r}, {\mathcal T}; \kappa, A_0, {\it \Delta}p)=aS_1 + \kappa S_2+ U_{\rm Fix}(A_0) - {\it \Delta}pV + U_{\rm S},  \nonumber \\
&& S_1=\sum_{ij} \left( {\bf r}_i-{\bf r}_j\right)^2, \quad S_2=\sum_{ij} (1-{\bf n}_i \cdot {\bf n}_j), 
\end{eqnarray} 
where $S_1$ is the Gaussian bond potential and $S_2$ is the bending energy with the bending rigidity $\kappa[1/k_BT]$; $k_B$ is the Boltzmann constant and $T$ is the temperature. We should note that ${\mathcal T}$ included in $S$ as a dynamical variable is only for the fluid model, and it is fixed for the fixed-connectivity model. The Hamiltonian $S$ is considered as the one for the $N$-particle system, and therefore $S$ can be viewed as the microscopic Hamiltonian. We should note that the coefficient $a$ of $S_1$, which is the microscopic surface tension, is fixed to $a\!=\!1$ hence force. The symbol $ {\bf n}_i$ denotes the unit normal vector of the triangle $i$, and $\sum_{ij}$ of $S_1$ and $S_2$ denote the sum over all nearest neighbor vertices and triangles, respectively. The potential $U_{\rm Fix}(A_0)$, which fixes the surface area $A$ to a constant $A_0$ without using the boundary $\Gamma$, is defined by
\begin{eqnarray} 
\label{Fix-potential-tension}
U_{\rm Fix}(A_0)= \left\{
     \begin{array}{@{\,}ll}
    \infty & \; (|1-A/A_0|>\epsilon_A=1/N_T),  \\
             0 & \; ({\rm otherwise}), 
     \end{array} 
               \right. 
\end{eqnarray}
where $N_T$ is the total number of triangles $N_T\!=\!2N\!-\!4$. The reason why the constraint is imposed on the real area $A$ instead of the projected area $A_p$ is because $A$ is well-defined while $A_p$ is not. Note also that the surface is allowed to have only in-plane deformation if $\epsilon_A$ is exactly zero, therefore $\epsilon_A$ must be a small positive number. The energy $-{\it \Delta}pV$ with the enclosed volume $V$ is well-defined only for the self-avoiding (SA) surfaces. Thus, the energy  $-{\it \Delta}pV$ should be included in the Hamiltonian together with the SA potential $U_{\rm S}$, which is defined by 
\begin{eqnarray} 
\label{SA-potential}
&&U_{\rm S}=\sum_{\it \Delta \Delta^\prime} U_{\rm S}({\it \Delta,\Delta^\prime}), \nonumber \\
&&U_{\rm S}({\it \Delta,\Delta^\prime})= \left\{
     \begin{array}{@{\,}ll}
    \infty & \; ({\rm triangles}\;{\it \Delta \Delta^\prime} \; {\rm intersect}), \\
             0 & \; ({\rm otherwise}), 
     \end{array} 
               \right.
\end{eqnarray}
where $\sum_{\it \Delta \Delta^\prime}$ denotes the sum over all non-nearest neighbor triangles.  This SA potential is considered as an extension of the one in the Doi-Edwards model for polymers \cite{Doi-Edwards-1986}, and it is slightly simpler as compared to the one in Ref. \cite{BOWICK-TRAVESSET-EPJE2001}.

The partition function $Z$ is defined by 
\begin{eqnarray} 
\label{micro-Z}
Z = \sum_{\mathcal T} \int^\prime \prod _{i=1}^{N} d {\bf r}_i \exp\left[-S({\bf r}, {\mathcal T})\right],
\end{eqnarray} 
where $\sum_{\mathcal T}$ denotes the sum over all possible triangulations. This  $\sum_{\mathcal T}$ is included in $Z$ only for the fluid surface model; the fixed-connectivity model is defined by $Z$ without $\sum_{\mathcal T}$. The prime in $ \int^\prime \prod _{i=1}^{N} d {\bf r}_i $ denotes that the center of mass of the surface is fixed at the origin of ${\bf R}^3$. We comment on the phase space measure in $Z$. The integration measure $\prod _{i=1}^{N} d {\bf r}_i$ can be replaced by $\prod _{i=1}^{N} d {\bf r}_i (q_i/3)^{\sigma}$, where $q_i$ is the coordination number of the vertex $i$ and $\sigma\!=\!3/2$ \cite{Gompper-Kroll-PRE1995,FDavid-NPB1985}. Since $q_i^\sigma$ can also be written as $q_i^{\sigma}\!=\!\exp(\sigma \log q_i)$, this measure effectively turns to be the term $-\sigma \sum_i\log q_i$ in the Hamiltonian. Therefore, this term is expected to influence the distribution of the coordination number on the fluid surface model. Therefore the fluid model in general depends on the integration measure which includes the coordination number. However, we assume $\sigma\!=\!0$ for simplicity in this paper just like in Refs. \cite{Gompper-Kroll-PRE1995,Koibuchi-Shobukhov-PHYSA2114}.

The surface tension $\gamma$ is mathematically understood in the context of HP model \cite{WHEATER-JP1994}. Indeed $\gamma$ is found from the scale invariance of $Z$, which is expressed by the equation $\partial_\alpha \log Z(\alpha{\bf r})|_{\alpha=1}\!=\!0 $. It is easy to see that $S_2$ and $U_{\rm S}$ are scale independent and $S_1(\alpha{\bf r})\!=\!\alpha^2 S_1({\bf r})$, ${\it \Delta}pV(\alpha{\bf r})\!=\!\alpha^3 {\it \Delta}pV({\bf r})$. The dependence of $S$ on the variable ${\mathcal T}$, which is necessary for the fluid surface model, is not explicitly written for simplicity. 
Since $Z(A_0;\alpha{\bf r})\!=\!Z(U_{\rm Fix}(A_0);\alpha{\bf r})\!=\!Z(U_{\rm Fix}(\alpha^{-2}A_0);{\bf r})\!=\!Z(\alpha^{-2}A_0;{\bf r})$, we see that the corresponding derivative $\partial_\alpha \log Z(A_0;\alpha{\bf r})|_{\alpha=1}$ can be written as $\partial_{\alpha^{-2}A_0} \log Z(\alpha^{-2}A_0;{\bf r})\partial_\alpha (\alpha^{-2}A_0)|_{\alpha=1}\!=\!-2A_0\partial_{A_0} \log Z(A_0;{\bf r})$ \cite{WHEATER-JP1994}. Thus, we have
\begin{eqnarray}
\label{surface-tension-org} 
&&\left(3(N-1)\alpha^{3(N-1)-1}Z(A_0;\alpha{\bf r}) \right. \nonumber \\
&& -2 \alpha^{3(N-1)+1}\sum _{\mathcal T}\int^\prime \prod _i d {\bf r}_i S_1 \exp\left[-S(\alpha{\bf r})\right] \nonumber \\ 
&& +3 \alpha^{3(N-1)+2}\sum _{\mathcal T}\int^\prime \prod _i d {\bf r}_i {\it \Delta}pV \exp\left[-S(\alpha{\bf r})\right]\nonumber \\
&&\left.\left.\left. -2A_0\alpha^{-3}\frac{\partial Z(\alpha^{-2}A_0;{\bf r})}{\partial A_0}\right) \right/ Z(A_0;\alpha{\bf r}) \right|_{\alpha=1}=0,
\end{eqnarray}
where $S(\alpha{\bf r})=\alpha^2S_1\!+\!\kappa S_2\!+\!U_{\rm Fix}(\alpha^{-2}A_0)\!-\!\alpha^3{\it \Delta}pV\!+\!U_S$, and $\sum _{\mathcal T}$ is included only for the fluid model. 
To evaluate the final term in Eq. (\ref{surface-tension-org}), we should recall that the free energy for the surface with the fixed boundary is given by $F\!=\!-\log Z\!=\!\tau A_p\!+\!\lambda H_{\rm el}\!-\!{\it \Delta}pV$ \cite{Cai-Lub-PNelson-JFrance1994}. Here, $k_BT$ ($k_B$ and $T$ are the Boltzmann constant and the temperature) is set to be $k_BT\!=\!1$,  the symbol $A_p$ denotes the projected area of the boundary, and $\lambda H_{\rm el}$ is the bending energy. However, the surface area $A_0$ is fixed while $A_p$ is unknown in the model of this paper, because the surface has no fixed  boundary, and therefore $A_p$ is replaced by the surface area $A_0$. Because of such replacement, the frame tension $\tau$ in this case should be changed to the surface tension $\gamma$.     
 Thus, we have
\begin{eqnarray}
\label{free-energy}
F=-\log Z(A_0,V)=\gamma A_0+\lambda H_{\rm el}-{\it \Delta}pV.
\end{eqnarray}
Note also that the bending energy term $\lambda H_{\rm el}$ does not contribute to the surface tension at least in the term \\$-2A_0\partial_{A_0} \log Z(A_0,V)$. Moreover, the term  $\lambda H_{\rm el}$ can be neglected from the Hamiltonian if the surface area and the volume become sufficiently large ($A_0\!\to\!\infty$, $V\!\to\!\infty$). In this case the terms $\gamma A_0$ and $ {\it \Delta}pV$ are apparently dominant because $\lambda H_{\rm el}$ is scale independent. For this reason, we do not go into detail of the term $\lambda H_{\rm el}$. Thus, from Eqs. (\ref{surface-tension-org}) and (\ref{free-energy}) we have 
\begin{eqnarray} 
\label{surface-tension}
\gamma|_{A}=(2\langle S_1\rangle_{A,{\it \Delta}p} - 3{\it \Delta}p\langle V\rangle_{A,{\it \Delta}p}-3N)/(2A_0),
\end{eqnarray} 
where  $\langle S_1\rangle_{A,{\it \Delta}p}$ and $\langle V\rangle_{A,{\it \Delta}p}$ are obtained by ACMCS.

Here, we should comment on the units of length and energy. Since the unit of $aS_1$ is $[k_BT]$ and the unit of $S_1$ is the length squared, we understand that the length unit is given by $\sqrt{k_BT/a}$, where $a$ is the microscopic surface tension. Note also that $a$ can be fixed to $a\!=\!1$ from the scale invariance of the partition function $Z$ in Eq. (\ref{micro-Z}). Indeed, $\int^\prime \prod _{i=1}^{N} d {\bf r}_i \exp\left[-S({\bf r}, {\mathcal T}; a, \kappa, A_0, {\it \Delta}p)\right]$ in $Z$ can be reduced to $\int^\prime \prod _{i=1}^{N} d {\bf r}_i^\prime \exp\left[-S({\bf r}^\prime, {\mathcal T}; 1, \kappa, A_0^\prime, {\it \Delta}p^\prime)\right]$ up to a multiplicative constant by the change of integration variable such that ${\bf r}\to{\bf r}^\prime\!=\!\sqrt{a}{\bf r}$ ($\Leftrightarrow {\bf r}\!=\!(1/\sqrt{a}){\bf r}^\prime$). This scale change does not influence the constant $A_0$, and therefore we introduce the new constant $A_0^\prime\!=\!aA_0$ for the scaled Hamiltonian. From this new $A_0^\prime$, it is easy to see $U_{\rm Fix}({\bf r}^\prime;A_0^\prime)\!=\!U_{\rm Fix}({\bf r};A_0)$ because of the relation that $A^\prime/A_0^\prime\!=\!A/A_0$ and the definition of $U_{\rm Fix}({\bf r};A_0)$ in Eq. (\ref{Fix-potential-tension}).  It is also easy to see that ${\it \Delta}p^\prime V^\prime\!=\!{\it \Delta}p V$, where ${\it \Delta}p^\prime$ is defined by ${\it \Delta}p^\prime\!=\!a^{-3/2}{\it \Delta}p$. Thus, using the fact that $S_2$ and $U_{\rm S}$ are scale independent, we prove  $S({\bf r}, {\mathcal T}; a, \kappa, A_0, {\it \Delta}p)\!=\!S({\bf r}^\prime, {\mathcal T}; 1, \kappa, A_0^\prime, {\it \Delta}p^\prime)$.

\subsection{Pressure difference ${\it \Delta}p$}
The pressure difference ${\it \Delta}p$ can also be calculated using almost the same model and procedure as those described above for $\gamma|_A$. Indeed, the constraint $U_{\rm Fix}(A_0)$ in Eq. (\ref{Disc-Eneg-tension}) can be replaced by $U_{\rm Fix}(V_0)$ to fix the enclosed volume $V$ to $V_0$; and the energy $ - {\it \Delta}pV$ should  be removed from the Hamiltonian.  In this model, ${\it \Delta}p$ is not an input parameter but it is produced as an output. Thus, the Hamiltonian is given by 
\begin{eqnarray}
\label{Disc-Eneg-dp} 
S({\bf r}, {\mathcal T}; \kappa, V_0)=S_1 + \kappa S_2+ U_{\rm Fix}(V_0) + U_{\rm S},  
\end{eqnarray} 
where  $U_{\rm Fix}(V_0)$ is defined by replacing $A/A_0$ with $V/V_0$ in Eq. (\ref{Fix-potential-tension})
\begin{eqnarray} 
\label{Fix-potential-dp}
U_{\rm Fix}(V_0)= \left\{
     \begin{array}{@{\,}ll}
    \infty & \; (|1-V/V_0|>\epsilon_V), \\
             0 & \; ({\rm otherwise}). 
     \end{array} 
               \right. 
\end{eqnarray}
In this expression, $\epsilon_V$ is defined in such a way that the enclosed volume $V$ satisfies $|V\!-\!V_0|\!<\!v_0$, where $v_0$ is the volume of the tetrahedron which consists of the regular triangles with the area $a_0\!=\!A_0/N_T$; $A_0\!=\!4\pi\left(3V_0/4\pi\right)^{2/3}$ is the area of the sphere of volume $V_0$. We should note that $a_0$ in VCMCS is not always identical with the mean triangle area $A/N_T$, because the area $A$ in VCMCS is not fixed to $A_0$. However, it is expected that  $A_0\!\simeq\!A$, because the surface always becomes a sphere in VCMCS although the surface fluctuations become large except for $\kappa\!\to\!\infty$ at least.

From the scale invariance of the partition function, we have
\begin{eqnarray} 
\label{pressure-difference}
{\it \Delta}p|_V=(3N-2\langle S_1\rangle_V)/(3V_0).
\end{eqnarray} 
Indeed, we see that $\partial_\alpha \log Z(\alpha^{-3}V_0;{\bf r})|_{\alpha=1}$ can be replaced by $-3V_0\partial_{V_0} \log Z(V_0;{\bf r})$ \cite{WHEATER-JP1994}. Thus, we have Eq. (\ref{pressure-difference}) by using the same procedure as described in the previous subsection. The symbol ${\it \Delta}p|_V$ is used for the calculated pressure to distinguish it from the input parameter ${\it\Delta}p$ in Eq. (\ref{Disc-Eneg-tension}).

It must be emphasized that ${\it \Delta}p|_V$ does not suffer from the problem encountered in the calculation of the surface tension caused by the difference between $A_p$ and $A$, because the energy increment due to the volume variation ${\it \Delta}V$ is independent of the way how the surface shape changes. On the contrary, the increment of the frame tension energy depends only on the change ${\it \Delta}A_p$ of the projected area $A_p$. This ${\it \Delta}A_p$ is not always identical to the change ${\it \Delta}A$ of the microscopic surface area $A$ of the sphere without the boundary.

\section{Monte Carlo Results}
The canonical Metropolis MC technique is used to update the variables ${\bf r}$ and $\mathcal T$, where $\mathcal T$ is updated only for the fluid surface model by using the standard bond flip technique.  
The acceptance rate $R_{\bf r}$ for ${\bf r}\to{\bf r}^\prime\!=\!{\bf r}\!+\!\delta {\bf r}$ can be controlled by tuning the radius $r_0$ of a small sphere, inside which  $\delta {\bf r}$ is chosen randomly. We fix $r_0$ to maintain $R_{\bf r}R_{\rm S}R_{\rm Fix}\!\simeq\!50\%$, where $R_{\rm S}$ and  $R_{\rm Fix}$ are the acceptance rates for the interactions $U_{\rm S}$,  $U_{\rm Fix}(A_0)$ or  $U_{\rm Fix}(V_0)$, respectively. The rate $R_{\rm Fix}$ is almost $100\%$ at sufficiently large $\kappa$ while it is  $70\%\sim 80\%$ in the limit of $\kappa\!\to\!0$. The fact that $R_{\rm Fix}$ is relatively high implies that the potentials $U_{\rm Fix}(A_0)$ and $U_{\rm Fix}(V_0)$ for the fixed area and the enclosed volume are suitable to the local update MC technique. The rate $R_{\rm S}$ also depends on $\kappa$ and  $R_{\rm S}\!\simeq\! 50\%$ or more for $\kappa\!\to\!0$ while $R_{\rm S}\!\simeq\! 100\%$ for $\kappa\!\to\!\infty$. All simulations are performed on the surface with $N\!=\!1442$. 
The total number of MCS is approximately $5\times10^6$ including $5\times10^5$ thermalization MCS.
\subsection{Dependence of $\gamma$ on the bending rigidity\label{gamma-1}}
\begin{figure}[tb]
\centering
\includegraphics[width=8.6cm]{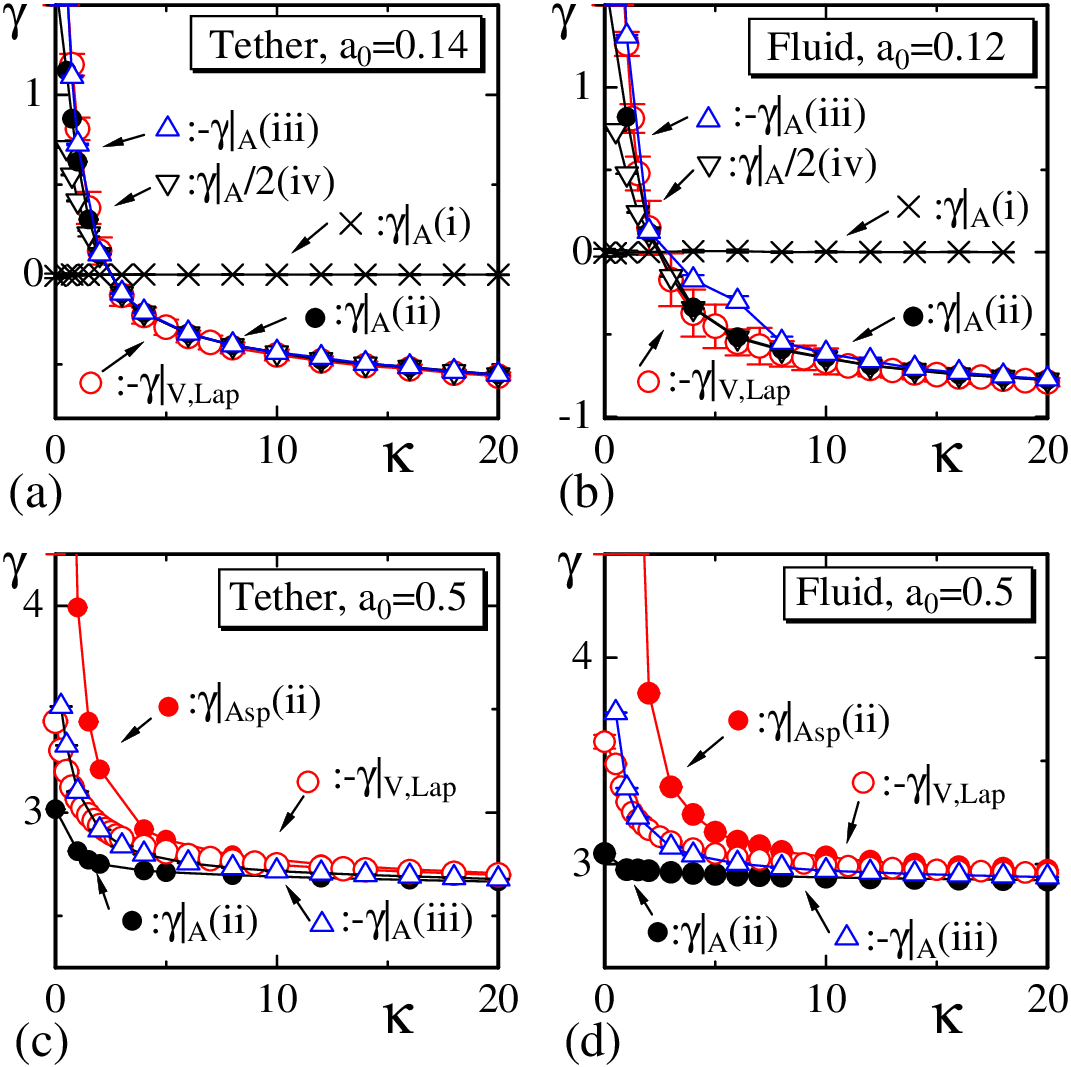}
\caption{(Color Online)  $\gamma$ vs. $\kappa$ at (a) $a_0(=\!A_0/N_T)\!=\!0.14$ (tethered), (b) $a_0\!=\!0.12$ (fluid), (c) $a_0\!=\!0.5$ (tethered) and (d) $a_0\!=\!0.5$ (fluid). The symbols ${\color{red}\bigcirc}$ ($\times$, $\bullet$, ${\color{red}\bullet}$, ${\color{blue}\bigtriangleup}$, $\bigtriangledown$) denote the results obtained by the constant volume (area) simulations. The lattice size is $N\!=\!1442$.} 
\label{fig-1}
\end{figure}

Let us check whether the MC results of $\gamma|_A$ are consistent with those of ${\it\Delta}p|_V$. We firstly perform VCMCS by varying $\kappa$ with constant $V_0$, which is fixed by using the triangle areas of the initial sphere so that the mean value $a_0\!=\!A_0/N_T$ becomes $a_0\!=\!0.14, \,0.5$ (tethered model) and $a_0\!=\!0.12,\, 0.5$ (fluid model). This $a_0$ approximately equals $a_0\!=\!0.216$ for the initial sphere of size $S_1/N\!=\!3/2$, which holds for the equilibrium surfaces without the potential $U_{\rm Fix}(A_0)$ under ${\it\Delta}p\!=\!0$. 
From the MC results ${\it\Delta}p|_V$ of the tethered and fluid models we obtain $\gamma|_{V,{\rm Lap}}$
using the Laplace formula 
\begin{eqnarray}
\label{Laplace-dp}
\gamma|_{V,{\rm Lap}}=(R/2){\it\Delta}p|_V,\quad R=\left(3V_0/4\pi\right)^{1/3}.
\end{eqnarray}
This $\gamma|_{V,{\rm Lap}}$ corresponds to the frame tension $\tau$. Here we use the symbol $\gamma|_{V,{\rm Lap}}$ for $\tau$  so that one can see how the frame tension is obtained.
These results are plotted (${\color{red}\bigcirc}$) in Figs. \ref{fig-1}(a)--(d). Note that the surface shape remains spherical in VCMCS  with constant $V_0$ even in the limit of $\kappa\!\to\!0$ for both tethered and fluid models.

 Next, we perform ACMCS with constant area $A_0$ and  ${\it\Delta}p$, both of which are supplied by the output of VCMCS that were performed to obtain the data denoted by the symbol (${\color{red}\bigcirc}$). Thus, the input parameters for ACMCS are given by 
\begin{eqnarray}
\label{inputs-for-A-const-1}
A_0=\langle A\rangle_V
\end{eqnarray}
and
\begin{eqnarray}
\label{inputs-for-A-const}
({\rm i})&& \;  {\it\Delta}p=-{\it\Delta}p|_V, \qquad
({\rm ii})  \;   {\it\Delta}p=0, \nonumber \\
({\rm iii})&& \;  {\it\Delta}p=-2{\it\Delta}p|_V, \quad \;({\rm iv})  \;{\it\Delta}p={\it\Delta}p|_V. 
\end{eqnarray}
Under these four different conditions in Eq. (\ref{inputs-for-A-const}) together with the one in Eq. (\ref{inputs-for-A-const-1}), we perform ACMCS and obtain the surface tensions $\gamma|_A$ from Eq. (\ref{surface-tension}). Thus we predict that the results $\gamma|_A(\kappa)$  satisfy
\begin{eqnarray}
\label{outputs-for-V-const}
({\rm i})&& \; \gamma |_A=0 (\times ), \qquad \quad
({\rm ii}) \; \gamma |_A=-\gamma|_{V,{\rm Lap}} (\bullet),  \\
({\rm iii})&& \; \gamma |_A=\gamma|_{V,{\rm Lap}} ({\color{blue}\bigtriangleup}), \; ({\rm iv}) \;  \gamma |_A=-2\gamma|_{V,{\rm Lap}} (\bigtriangledown)\nonumber
\end{eqnarray}
corresponding to the conditions in Eq. (\ref{inputs-for-A-const}), if the surface remains spherical (the data are plotted in Fig. \ref{fig-1}). 

The predictions in Eq. (\ref{outputs-for-V-const}) come from the expectations
\begin{eqnarray}
\label{expectation-1}
V_0=\langle V\rangle_{A,{\it \Delta}p},\quad \langle S_1\rangle_V=\langle S_1\rangle_{A,{\it \Delta}p}.
\end{eqnarray}
 It is easy to prove the predictions in Eq. (\ref{outputs-for-V-const}) using  Eq. (\ref{expectation-1}). 
 Firstly, performing VCMCS with constant $V_0$ we have ${\it \Delta}p|_V=(3N-2\langle S_1\rangle_V)/(3V_0)$ by Eq. (\ref{pressure-difference}). Next, the condition (iii) $ {\it\Delta}p=-2{\it\Delta}p|_V$ is, for example, used as an input as well as the constant area $A_0=\langle A\rangle_V$ for ACMCS. Thus, we have from Eq. (\ref{surface-tension}) that 
\begin{eqnarray}
\gamma|_{A} &=& \left(2\langle S_1\rangle_{A,{\it \Delta}p} + 6{\it \Delta}p|_V\langle V\rangle_{A,{\it \Delta}p}-3N\right)/(2\langle A\rangle_V) \nonumber \\
&=&\frac{1}{2\langle A\rangle_V}\left(2\langle S_1\rangle_{A,{\it \Delta}p} + 6\frac{3N-2\langle S_1\rangle_V}{3V_0} \langle V\rangle_{A,{\it \Delta}p}-3N\right) \nonumber \\
&=&\left(3N-2\langle S_1\rangle_V\right) /(2\langle A\rangle_V)\nonumber \\
&=&(\bar R/2)\left(3N-2\langle S_1\rangle_V\right) /(3V_0)=(R/2){\it\Delta}p|_V,
\end{eqnarray}
where the relations in Eq. (\ref{expectation-1}) are used to obtain the third equality. 
In the forth equality, $\bar R$ is defined by $\bar R\!=\!(3\langle V\rangle_{A,{\it \Delta}p})/\langle A\rangle_V$ which equals $(3V_0)/\langle A\rangle_V$ because of Eq. (\ref{expectation-1}). This $\bar R$ can be identified with $R\!=\!(3V_0/4\pi)^{1/3}$ in Eq. (\ref{Laplace-dp}), because the surface in VCMCS is expected to be a sphere. Thus, we have (iii) in Eq. (\ref{outputs-for-V-const}). The other formulas in Eq. (\ref{outputs-for-V-const}) are also obtained in the same way. The basic assumption for Eq. (\ref{outputs-for-V-const}) is that the surface must be a sphere in both ACMCS and VCMCS.

These equations in Eq. (\ref{expectation-1}) are almost trivial because the enclosed volume $V_0$ is uniquely determined by its surface area $A_0$ if the surface is a complete sphere. Thus we understand that the surface tension $\gamma|_A$ is consistent to the frame tension $\gamma|_{V,{\rm Lap}}$ when the surface is a smooth sphere (in the limit of $\kappa\to \infty$). This fact is in good agreement with the result of Ref. \cite{Fournier-PRL2008} that the deviation between $\tau$ and $\gamma$ is  proportional to the temperature $T$ (at least $T\!\to\!  0$), because  $\kappa\propto 1/k_BT$ here in this paper. 

Here we show the meanings of the data in Fig. \ref{fig-1} in more detail.  Firstly, we note that the effective frame tension $\gamma|_{V,{\rm Lap}}$ (${\color {red}\bigcirc}$) obtained by VCMCS and plotted in Fig. \ref{fig-1} (a) has $\gamma|_{V,{\rm Lap}}\!<\!0$ ($\gamma|_{V,{\rm Lap}}\!>\!0$) for $\kappa\!<\!3$ ($\kappa\!>\!3$), because  $-\gamma|_{V,{\rm Lap}}$ (minus sign) is plotted in the figures. Because of the same reason, the data $\gamma |_A$ (${\color{blue}\bigtriangleup}$) obtained by ACMCS and plotted in Fig. \ref{fig-1} (a)  has also  $\gamma|_A\!<\!0$ ($\gamma|_A\!>\!0$) for $\kappa\!<\!3$ ($\kappa\!>\!3$). We can also see that ${\it\Delta}p\!=\!-2{\it\Delta}p|_V$, which is the one of the input data corresponding to (iii) in Eq. (\ref{inputs-for-A-const}) for this ACMCS, has ${\it\Delta}p\!>\!0$ (${\it\Delta}p\!<\!0$) for $\kappa\!<\!3$ ($\kappa\!>\!3$) because of Eq. (\ref{Laplace-dp}). We see from Fig. \ref{fig-1} (a) that $-\gamma |_A\!=\!-\gamma|_{V,{\rm Lap}}$, which is just the expected (iii) $\gamma |_A\!=\!\gamma|_{V,{\rm Lap}}$ in Eq. (\ref{outputs-for-V-const}), is satisfied except for the region for $\kappa\!\to\!0$. @

Figure \ref{fig-1} (b) shows that the graph of (iii) $\gamma |_A\!=\!\gamma|_{V,{\rm Lap}} ({\color{blue}\bigtriangleup})$ is slightly broken in  the region $\kappa\!\simeq\!5$. The reason of this deviation is that the surface shape becomes prolate \cite{Gompper-Kroll-PRE1995,Koibuchi-Shobukhov-PHYSA2114} under relatively large negative pressure ${\it\Delta}p\!=\!-2{\it\Delta}p|_V$ at $\kappa\!\simeq\!5$. In the region of small $\kappa$, such as $\kappa<3$,  the pressure difference $-2{\it\Delta}p|_V$, which is used as an input ${\it\Delta}p$ of ACMCS, is positive, and hence the surface becomes a sphere in ACMCS. In the region of sufficiently large $\kappa$ the surface also becomes a sphere even though ${\it\Delta}p\!=\!-2{\it\Delta}p|_V$ is negative. Only for the intermediate values of $\kappa$ the surface happens to be the prolate because of the negative input of ${\it\Delta}p$. 

We also see a break in the prediction (ii) of Eq. (\ref{outputs-for-V-const}) for the region of small $\kappa$ in Figs. \ref{fig-1}(c),(d). It comes only from the deviation of $A_0$ from $A_p$, which is unknown, but expected in that region. Indeed, $\gamma|_{\rm A}$ ($\gamma|_{\rm A_{\rm sp}}$) is slightly smaller (larger) than $-\gamma|_{V, {\rm Lap}}$, where $\gamma|_{\rm A_{\rm sp}}$ is calculated by replacing $A_0$ with  $A_{\rm sp}\!=\!4\pi\left(3\langle V\rangle_{A,{\it \Delta}p}/4\pi\right)^{2/3}$ in Eq.  (\ref{surface-tension}). The value of $A_p$ is expected to satisfy $A_{\rm sp}\!<\!A_p\!<\!A_0$ for the condition (ii). Note that the results ${\it\Delta}p|_V$ of VCMCS become large negative at $a_0\!=\!0.5$ than those obtained at $a_0\!=\!0.14$ and $a_0\!=\!0.12$. The large negative input ${\it\Delta}p(=\!{\it\Delta}p|_V\!<\!0)$ makes the surface stomatocyte  \cite{Gompper-Kroll-PRE1995,Koibuchi-Shobukhov-PHYSA2114} in the ACMCS for $a_0\!=\!0.5$, and therefore the expectation (iv) $\gamma |_A/2\!=\!-\gamma|_{V,{\rm Lap}}$ is actually not satisfied in both tethered and fluid surfaces in this case. These deviations come from the fact that the predictions in Eqs. (\ref{outputs-for-V-const}) are only satisfied for a sphere not only in VCMCS but also in ACMCS. Note also that the prediction (i) of Eq. (\ref{outputs-for-V-const}), which is not depicted, is correct also in the whole region of $\kappa$ including $\kappa\!=\!0$ for $a_0\!=\!0.5$. 

\begin{figure}[tb]
\centering
\includegraphics[width=8.6cm]{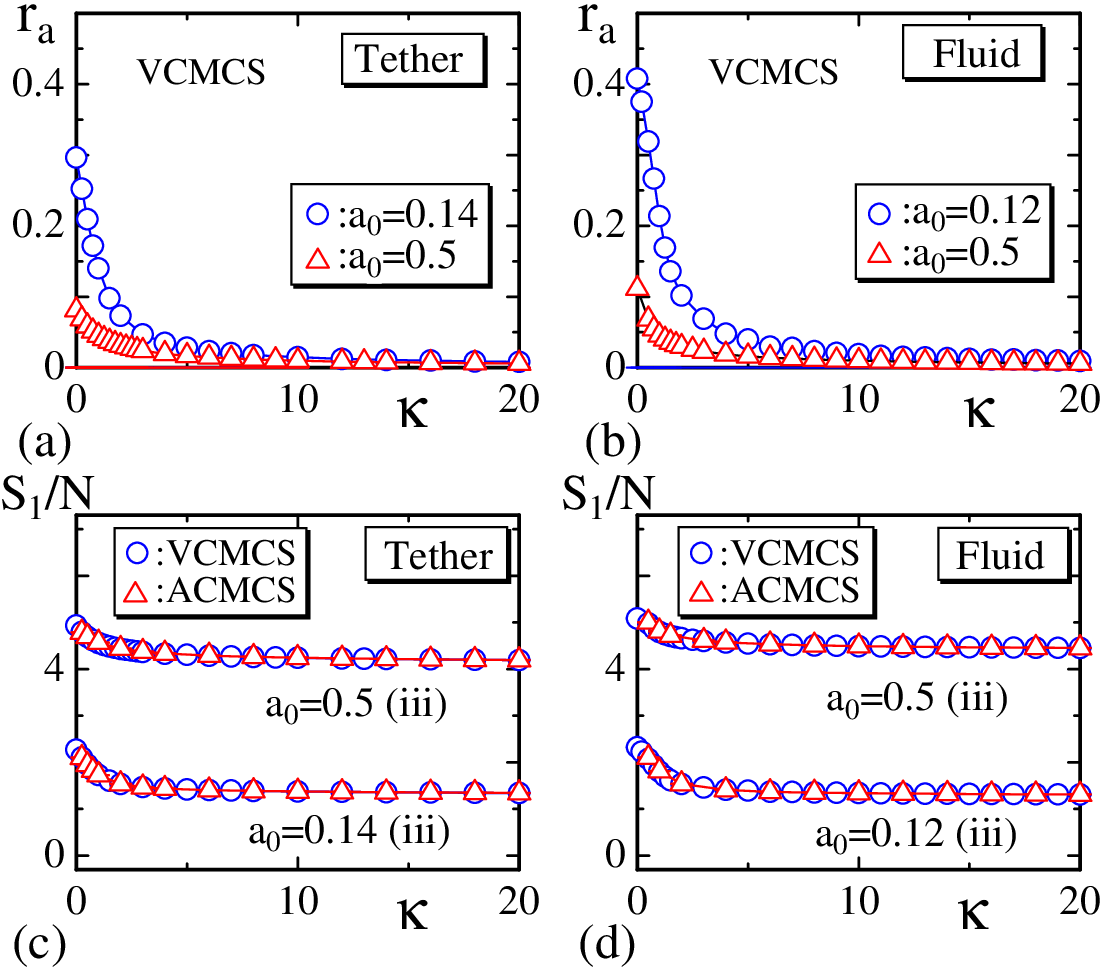}
\caption{(Color Online)  The ratio $r_a\!=\!(\langle A\rangle_V/4\pi)^{1/2}/(3V_0/4\pi)^{1/3}\!-\!1$ 
vs. $\kappa$ for (a) the tethered and (b) the fluid models.  The Gaussian bond potential $S_1/N$ vs. $\kappa$ for the check of the second equation in Eq. (\ref{expectation-1}) for (c) the tethered and (d) the fluid models. The symbol $a_0\!=\!0.5$(iii) in (c) denotes that both VCMCS and ACMCS are performed with $a_0\!=\!0.5$, and ACMCS is performed under the condition (iii) in Eq. (\ref{inputs-for-A-const}).  The lattice size is $N\!=\!1442$. }
\label{fig-2}
\end{figure}
To show how the surface of VCMCS deviates from a complete sphere, we plot the ratio defined by \\ $r_a\!=\!(\langle A\rangle_V/4\pi)^{1/2}/(3V_0/4\pi)^{1/3}\!-\!1$  in Figs. \ref{fig-2}(a),(b), where  $(\langle A\rangle_V/4\pi)^{1/2}$ is the mean radius obtained from the triangle area and $(3V_0/4\pi)^{1/3}$ is the one from the enclosed volume. We find that $(\langle A\rangle_V/4\pi)^{1/2}$ deviates from $(3V_0/4\pi)^{1/3}$ only in the region of small $\kappa$ both in the tethered and fluid surfaces. The reason of the deviation comes from the large surface fluctuations expected at $\kappa\!\to\! 0$.

The second equation in Eq. (\ref{expectation-1}) can also be checked (Figs. \ref{fig-2}(c),(d)). The symbol $a_0\!=\!0.5$(iii) in Fig. \ref{fig-2}(c) denotes that both VCMCS and ACMCS are performed with $a_0\!=\!0.5$, and ACMCS is performed under the condition (iii) in Eq. (\ref{inputs-for-A-const}). The Gaussian bond potential $\langle S_1\rangle_V$ obtained by VCMCS is in good agreement with $\langle S_1\rangle_{A,{\it \Delta}p}$ both in the tethered and the fluid surfaces except at $\kappa\!\to\!0$. This equation in Eq. (\ref{expectation-1}) holds also for the other three conditions in Eq. (\ref{inputs-for-A-const}) almost in the whole region of $\kappa$ just like in Figs. \ref{fig-2}(c),(d). Since the surface is not a complete sphere at sufficiently small $\kappa$ due to the surface fluctuations, the equations in Eq. (\ref{expectation-1}) are not always trivial. This leads us to understand that the second of Eq. (\ref{expectation-1}) implies that the mapping ${\bf r}:M\to {\bf R}^3$ under the constraint of constant area mathematically shares a common property with the one under the constraint of constant volume, as mentioned in the introduction. In this sense, the simulations for the surface/frame tension in this paper can also serve as a check for this property of the constrained mappings.

\begin{figure}[tb]
\centering
\includegraphics[width=7.0cm]{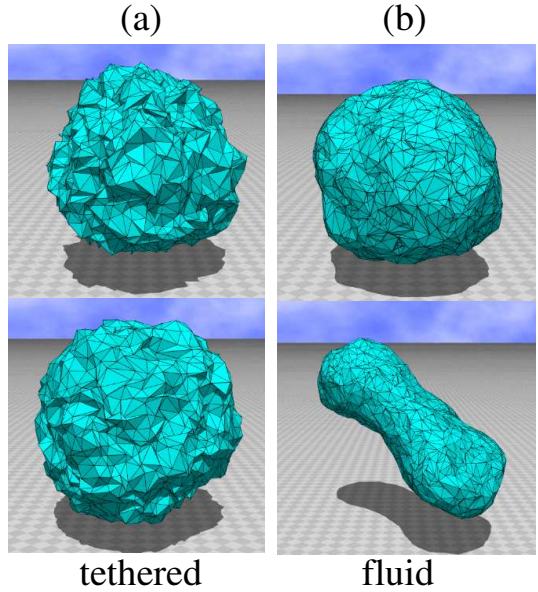}
\caption{(Color Online) Snapshots of the fixed-connectivity and the fluid surfaces obtained in VCMCS (upper) and ACMCS (lower) at (a) $a_0\!=\!0.14$, $\kappa\!=\!0.5$, (b) $a_0\!=\!0.12$, $\kappa\!=\!6$. The scales of the figures are different from each other.} 
\label{fig-3}
\end{figure}
Snapshots in Figs. \ref{fig-3}(a),(b) correspond to the data in Fig. \ref{fig-1} for the (a) tethered and (b) fluid surfaces. The surfaces in the upper (lower) row are those obtained in VCMCS (ACMCS). In ACMCS, ${\it\Delta}p$ is fixed to (a) ${\it\Delta}p\!=\!1.14$ and (b) ${\it\Delta}p\!=\!-0.418$, both of which correspond to (iii) in Eq. (\ref{inputs-for-A-const}). A small deviation in (iii) of Eq. (\ref{outputs-for-V-const}) seen in Fig. \ref{fig-1}(b) comes from the fact that the surface shape is not spherical but prolate (see Fig. \ref{fig-3}(b)) as mentioned above. The surfaces in Figs. \ref{fig-3}(a) are those obtained from the calculations of $\gamma|_A$ and ${\it\Delta}p$ for the checks in Eq. (\ref{outputs-for-V-const}), where no inconsistency was found for large $\kappa$ at least. 

\subsection{Dependence of $\gamma$ on the area\label{gamma-2}}
\begin{figure}[tb]
\centering
\includegraphics[width=8.6cm]{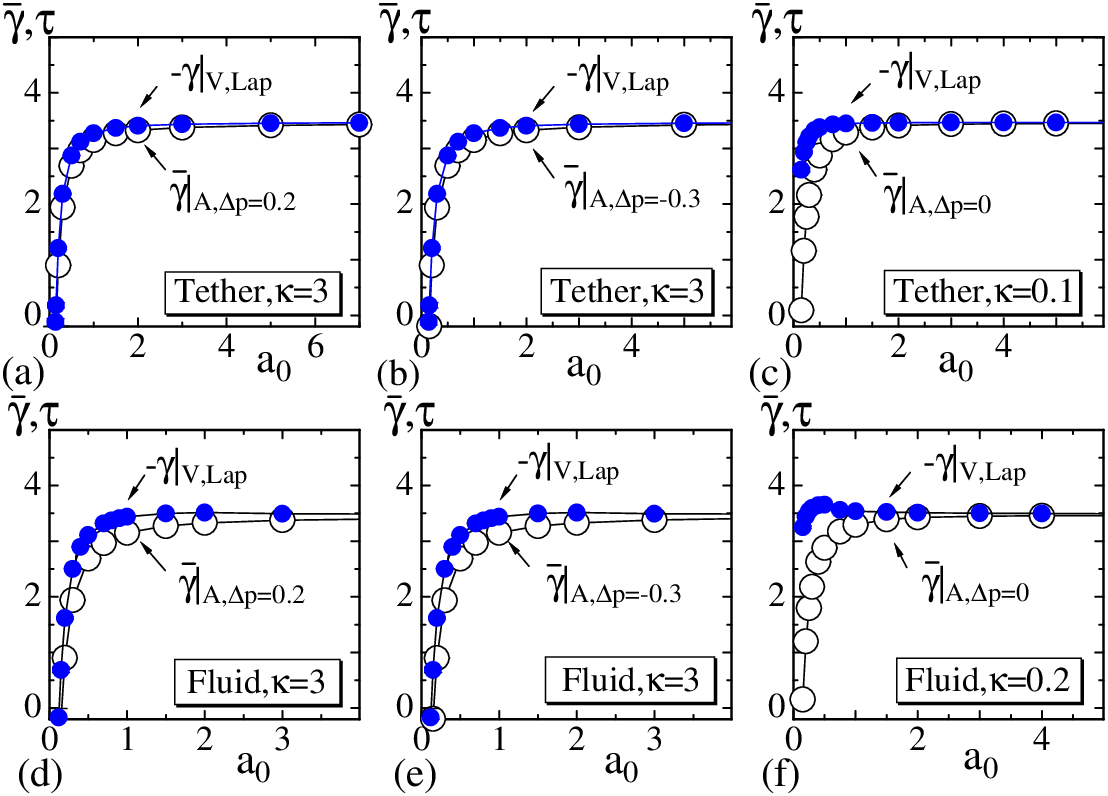}
\caption{(Color Online) $-\gamma|_{V,{\rm Lap}}$ (${\color{blue} \bullet}$) and ${\bar \gamma}|_{A,{\it\Delta}p}$ ($\bigcirc$) vs. $a_0(=A_0/N_T)$ for the tethered  and fluid models.   ${\it\Delta}p\!=\!0.2, -0.3, 0$ are assumed for the constant area simulations.  The lattice size is  $N\!=\!1442$.} 
\label{fig-4}
\end{figure}
Now we proceed to an additional check for the results of $\gamma|_A$ and $\gamma|_{V,{\rm Lap}}$ obtained in the ACMCS and VCMCS, respectively, by varying $A_0/N_T(=\!a_0)$ with fixed $\kappa$. In contrast to the first check described above, these two different simulations can be performed independently. The parameter $a_0$ is exactly identical to the mean value of the triangle area $A/N_T$ for ACMCS, while $a_0$ is only an input parameter to fix $V_0$ but almost identical to $A/N_T$ in VCMCS in the large $\kappa$ region at least as shwon in Figs. \ref{fig-2}(a),(b). Thus we obtain ${\it\Delta}p|_V$, from which $\gamma|_{V,{\rm Lap}}$ is calculated by means of the Laplace formula in Eq. (\ref{Laplace-dp}). Therefore we expect that the simulation results shown in Figs. \ref{fig-4}(a)--(f) satisfy
\begin{eqnarray}
\label{gamma-vs-A}
-\gamma|_{V,{\rm Lap}} ({\color{blue}\bullet})={\bar \gamma}|_{A,{\it\Delta}p} (\bigcirc),
\end{eqnarray}
where 
${\bar \gamma}|_{A,{\it\Delta}p}$ is defined by
\begin{eqnarray}
\label{gamma_bar}
{\bar \gamma}|_{A,{\it\Delta}p}=\left(2\langle S_1\rangle_{A,{\it\Delta}p}-3N\right)/(2A_0).
\end{eqnarray}

The expectation in Eq. (\ref{gamma-vs-A}) can be obtained as follows: The left hand side of Eq. (\ref{gamma-vs-A}) is given by using the Laplace formula in Eq. (\ref{Laplace-dp}) such that
\begin{eqnarray}
\label{B2}
-\gamma|_{V,{\rm Lap}}&=&-(R/2){\it\Delta}p|_V\nonumber \\
&=& -(R/2)\left(3N-2\langle S_1\rangle_V\right)/(3V_0)  \nonumber \\
&=& -\left(3N-2\langle S_1\rangle_V\right)/(2A_0),
\end{eqnarray}
where the second equality comes from Eq. (\ref{pressure-difference}), and the final equality is derived from the relations $R/(3V_0)\!=\!1/A_0$ and $4\pi R^2\!=\!A_0$. These relations are satisfied  because VCMCS and ACMCS are respectively performed with the constants $V_0$ and $A_0$, which satisfy $A_0\!=\!4\pi\left(3V_0/4\pi\right)^{2/3}$. The final expression in Eq. (\ref{B2}) is identified with ${\bar \gamma}|_{A,{\it\Delta}p}$ in Eq. (\ref{gamma_bar}) using the assumed equation 
\begin{eqnarray}
\label{expectation-2}
 \langle S_1\rangle_{A,{\it\Delta}p}=\langle S_1\rangle_V,
\end{eqnarray}
which is understood as a function of $a_0$. 
This is also a reasonable relation expected for the Gaussian bond potentials obtained in both simulations at least for sufficiently large $a_0$. The relation in  Eq. (\ref{expectation-2}) is weaker than those of Eq. (\ref{expectation-1}) in the sense that the surface shape of ACMCS is not necessarily identical to the one of VCMCS: stomatocyte, cup-like and dumbbell, even branched-polymer, are allowed in ACMCS for the check of Eq. (\ref{gamma-vs-A}). The basic assumption for Eq. (\ref{gamma-vs-A}) is that the surface must be a sphere only in VCMCS. 
 We note that the surface shape of the latter simulation (=VCMCS) must be spherical because the Laplace formula for a sphere is assumed to yield $\gamma|_{V,{\rm Lap}}$. This requirement is always fulfilled as mentioned above. Note also that the constant value of $\gamma|_{V,{\rm Lap}}$ in the limit of $a_0\!\to\!\infty$ is independent of whether the model is tethered or fluid.

The value ${\bar \gamma}|_{A,{\it\Delta}p}$ is identical to the surface tension in Eq. (\ref{surface-tension}) if ${\it\Delta}p\!=\!0$. However, it must be emphasized that the relation in Eq. (\ref{gamma-vs-A}) does not always imply that the surface tension $\gamma|_{A,{\it\Delta}p}$ can be identified with the frame tension  $\gamma|_{V,{\rm Lap}}$ but it only implies that ${\bar \gamma}|_{A,{\it\Delta}p}$ is equal to the minus of $\gamma|_{V,{\rm Lap}}$ obtained under the assumed condition.

An additional comment on the results in Fig. \ref{fig-4} is as follows: The negative $\gamma|_{V,{\rm Lap}}$ comes from the negative output ${\it\Delta}p|_V$ obtained in VCMCS (see Eq. (\ref{Laplace-dp})). In other words, this negative pressure is understood to be the one caused by the negative frame tension at sufficiently large $a_0$. This implies that the pressure difference becomes zero if the positive pressure ${\it\Delta}p\!=\!-{\it\Delta}p|_V$ can be externally supplied. Therefore, the results in Fig. \ref{fig-4} also imply that the frame tension in the limit of $a_0\!\to\!\infty$ becomes independent of $a_0$ under a suitable external pressure  ${\it\Delta}p(\propto \!1/R)$ for spherical surfaces, and that this property is also independent of whether the frame tension is negative or not. 
 
\begin{figure}[tb]
\centering
\includegraphics[width=8.6cm]{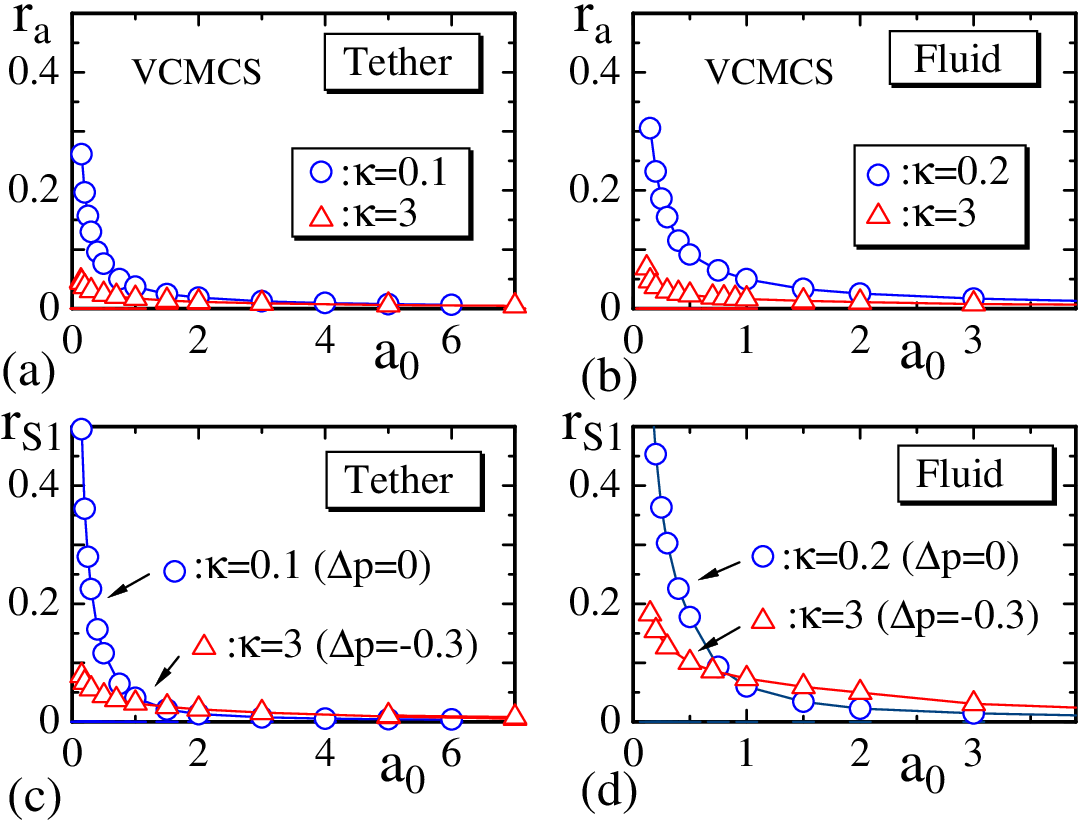}
\caption{(Color Online)  The ratio $r_a\!=\!(\langle A\rangle_V/4\pi)^{1/2}/(3V_0/4\pi)^{1/3}\!-\!1$  vs. $a_0$ for (a) the tethered and (b) the fluid models in VCMCS. The ratio $r_{S_1}\!=\!|1-\langle S_1\rangle_V/\langle S_1\rangle_{A,{\it\Delta}p}|$ for (c) the tethered and (d) the fluid models. The symbol $\kappa\!=\!0.1({\it\Delta}p\!=\!0)$ in (c) denotes that both VCMCS and ACMCS are performed with $\kappa\!=\!0.1$, and ACMCS is performed under ${\it\Delta}p\!=\!0$. The lattice size is $N\!=\!1442$.  }
\label{fig-5}
\end{figure}
To show how the surface of VCMCS deviates from a complete sphere, we plot the ratio \\ $r_a\!=\!(\langle A\rangle_V/4\pi)^{1/2}/(3V_0/4\pi)^{1/3}\!-\!1$ in Figs. \ref{fig-5}(a),(b). We find that $(\langle A\rangle_V/4\pi)^{1/2}$ is almost identical to $(3V_0/4\pi)^{1/3}$ for large $a_0$ region. To the contrary, in the limit of $a_0\!\to\!0$ the deviations of $(\langle A\rangle_V/4\pi)^{1/2}$ from $(3V_0/4\pi)^{1/3}$ are relatively large for the region of small $\kappa$. This is also expected from the results shown in Figs. \ref{fig-2}(a),(b). 

The equation (\ref{expectation-2}) can be checked by the ratio $r_{S_1}\!=\!|1-\langle S_1\rangle_V/\langle S_1\rangle_{A,{\it\Delta}p}|$, which is plotted in Figs. \ref{fig-5}(c),(d). The value of $S_1/N$ becomes very large in the limit of $a_0\!\to\!\infty$, and for this reason the ratio $r_{S_1}$ is plotted instead of $S_1/N$. The symbol $\kappa\!=\!0.1({\it\Delta}p\!=\!0)$ in Fig. \ref{fig-5}(c) denotes that both VCMCS and ACMCS are performed with $\kappa\!=\!0.1$, and ACMCS is performed under ${\it\Delta}p\!=\!0$. The data in Fig. \ref{fig-5}(c) correspond to those in Figs. \ref{fig-4}(b),(c) for the tethered model, while the data in Fig. \ref{fig-5}(d) correspond to those in Figs. \ref{fig-4}(e),(f) for the fluid model. We see that $r_{S_1}\!\to\!0$ $(a_0\!\to\!\infty)$ and therefore the relation in Eq. (\ref{expectation-2}) is satisfied at sufficiently large $a_0$ in both models.

\begin{figure}[tb]
\centering
\includegraphics[width=7.0cm]{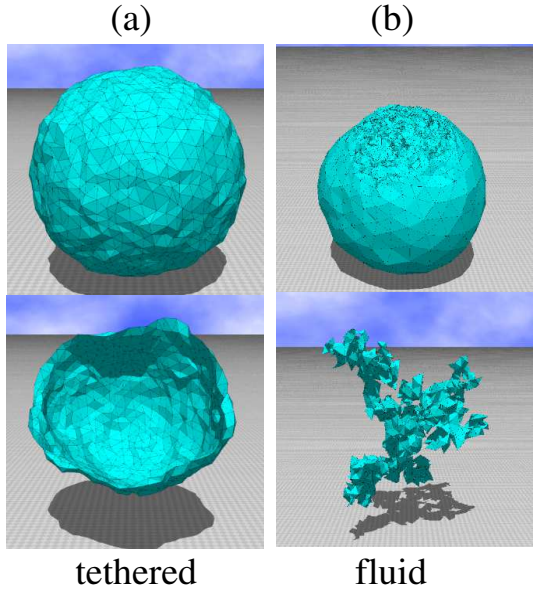}
\caption{(Color Online) Snapshots of the fixed-connectivity and fluid surfaces obtained in VCMCS (upper) and ACMCS (lower)  (a) $a_0\!=\!1$, $\kappa\!=\!3$, and (b) $a_0\!=\!4$, $\kappa\!=\!0.2$. The scales of the figures are different.} 
\label{fig-6}
\end{figure}
Snapshots in Figs. \ref{fig-6}(a),(b) correspond to  data presented in Fig. \ref{fig-4} (c) and Fig. \ref{fig-4} (f). In ACMCS (lower), ${\it\Delta}p$ is fixed to ${\it\Delta}p\!=\!-0.3$ (Fig. \ref{fig-6}(a)) and ${\it\Delta}p\!=\!0$ (Fig. \ref{fig-6}(b)).  The surfaces in Figs. \ref{fig-6}(a), (b) are obtained from the simulations for the calculations of $\gamma$ and ${\it\Delta}p$ for the check in Eq. (\ref{gamma-vs-A}). No deviation was found in the data from VCMCS and ACMCS for large $a_0$ at least. In Figs. \ref{fig-6}(a), (b) the surfaces obtained in ACMCS (Fig. \ref{fig-6}(b) lower)  differ from the sphere, however the relation in Eq.(\ref{expectation-2}) is satisfied and for this reason $\gamma$ is considered to be calculated consistently with ${\it\Delta}p$ according to the expectation in Eq. (\ref{gamma-vs-A}), although the configurations of the fluid model obtained in VCMCS at large $a_0$ are fluid only partly (Fig. \ref{fig-6}(b) upper).

Note that the acceptance rate of the bond flip $R_{\mathcal T}$ is very low ($R_{\mathcal T}\leq 0.05$) in VCMCS for the fluid model at $a_0 (>\!0.75)$. The reason for this is such a large energy change as ${\it \Delta}S_1$, which is caused by the bond flip; consequently the vertex diffusion is localized. For this reason, the surface is not completely fluid but - only partly - fluid for large $a_0$ in VCMCS for the fluid model. As a consequence the density of vertices becomes non-uniform on the surface, and therefore the surfaces at large $a_0$ may be different from the ordinary fluid ones at relatively small $a_0$. Indeed, we can see in the snapshot in Fig. \ref{fig-6}(b) that the density is higher in the upper side of the sphere than in the lower side. The lower side of the sphere is  relatively smooth and is composed of large size triangles. Almost all coordination numbers $q$ are expected to be $q\!=\!6$. It implies that there is no vertex diffusion in the lower side. Thus, the localization of vertex diffusion is expected on the surface in Fig. \ref{fig-6}(b).     

To avoid the localization of vertices, we can introduce the potential $U_2\!=\!\sum_i U_{2i}(a_0)$ with
\begin{eqnarray} 
\label{Fix-triangle-area}
U_{2i}(a_0)= \left\{
     \begin{array}{@{\,}ll}
    \infty & \; (|1-a_i/a_0|>\epsilon_2), \\
             0 & \; ({\rm otherwise}), 
     \end{array} 
               \right.
\end{eqnarray}
where $a_i$ is the area of triangle $i$ and $\epsilon_2\!=\!0.5$ for example. Due to this potential, every triangle area $a_i$ follows $(1/2)a_0 \!\leq\! a_i \!\leq\! (3/2)a_0$ for $\epsilon_2\!=\!0.5$. As a consequence, the total area $A$ is also constrained to obey the same relation as $a_i$. Recalling that the surface shape is always spherical in VCMCS, we find that the total surface area in VCMCS is limited only by the constant volume $V_0$ and hence it is not influenced by $U_2$. Therefore the formula for ${\it \Delta}p|_V$ in Eq.(\ref{pressure-difference}) is not influenced by $U_2$ as well. The potential $U_2$ has also no influence on the formula for $\gamma_A$ in Eq.(\ref{surface-tension}), because the area $A$ is fixed to $A_0$ by $U_{\rm Fix}(A_0)$ in Eq. (\ref{Fix-potential-tension}) in ACMCS. Thus, we can check whether or not $\gamma_A$ is consistent with ${\it \Delta}p|_V$ using the configurations in which the vertices are prevented from localization by the potential $U_2$. Indeed, the results in Fig. \ref{fig-4}(f) remain almost unchanged or relatively close to those in Figs. \ref{fig-4}(d) and \ref{fig-4}(e) if the potential $U_2$ is introduced in both ACMCS and VCMCS. In the ACMCS with $U_2$ the surface configurations are the branched polymer; they are identical to those without $U_2$, while in the VCMCS with $U_2$  the vertices freely diffuse over the surface even at relatively large $a_0$ and hence the configurations are manifestly different from those without $U_2$.

\section{Summary and conclusions}
In this paper, we present the surface tension $\gamma$ and the pressure difference ${\it\Delta}p$ of spherical membranes together with the calculation techniques on triangulated surfaces for the discretization of the canonical surface model of Helfrich and Polyakov. Using the real area $A$ of spherical surface, $\gamma$ is calculated by means of the area constant MC simulations (ACMCS). Using the Laplace formula, we evaluate the effective frame tension $\gamma|_{V,{\rm Lap}}$ from ${\it\Delta}p$, which is also calculated by means of the volume constant MC simulations (VCMCS). Thus, we have shown that the surface tension $\gamma$, conjugate to the real surface area $A$, is consistent with $\gamma|_{V,{\rm Lap}}$ at sufficiently large  $\kappa$ or $a_0(=A_0/N_T)$. Our results also show that this consistency holds for those data obtained on non-spherical surfaces in ACMCS at sufficiently large $a_0$. Thus, the results shown in this paper support that the frame tension can be evaluated by $\gamma$ by means of ACMCS with the proper input data for the constant area and the pressure difference under suitable condition such as sufficiently large $\kappa$ or $a_0$.

\begin{acknowledgements}
 This work is supported in part by the Grant-in-Aid for Scientific Research (C) Number 26390138. We acknowledge the support of the Promotion of Joint Research 2014, Toyohashi University of Technology. We are grateful to K. Osari and S. Usui for the computer analyses. 
\end{acknowledgements}



\end{document}